\documentclass[%
 reprint,
nofootinbib,
 amsmath,amssymb,
 aps,
prb,
longbibliography
]{revtex4-2}

\allowdisplaybreaks

\usepackage[utf8]{inputenc}
\usepackage{natbib}
\usepackage{mathtools}
\usepackage{graphicx}
\usepackage{dcolumn}
\usepackage{bm}

\usepackage{cancel}

\usepackage[dvipsnames]{xcolor}

\usepackage{graphicx}
\usepackage{soul}      
\usepackage{color}
\usepackage[colorlinks, citecolor=blue,linkcolor=blue,urlcolor=blue]{hyperref}
\usepackage{verbatim}
\usepackage{multirow}
\usepackage{bbold}
\usepackage{caption}
\usepackage{subcaption}
\usepackage{hyperref}
\hypersetup{
           breaklinks=true,   
           colorlinks=true,   
           pdfusetitle=true,  
           pdftitle = {Whence NCFT?}                   
        }

 \newcommand{\beq}{\begin{equation}}
 \newcommand{\eeq}{\end{equation}}
 \newcommand{\bea}{\begin{eqnarray}}
 \newcommand{\eea}{\end{eqnarray}}

 \newcommand{\half}{\frac{1}{2}}
 \newcommand{\bary}{\begin{array}}
 \newcommand{\eary}{\end{array}}
 
 \newcommand{\changes}[1]{{#1}}
 
  \newcommand{\bra}[1]{{{\langle #1 |}}}
 \newcommand{\ket}[1]{{| #1 \rangle}}
 \newcommand{\braket}[2]{\langle #1 \vert #2 \rangle}

  \newcommand{\Bra}[1]{{{\langle\langle #1 |}}}
 \newcommand{\Ket}[1]{{| #1 \rangle\rangle}}
 \newcommand{\Braket}[2]{\langle\langle #1 | #2 \rangle \rangle}
 
\newcommand{\term}[1]{\left( #1 \right)}
\newcommand{\abs}[1]{\left| #1 \right|}
\newcommand{\comm}[1]{\left[ #1 \right]}
\newcommand{\curly}[1]{\left\{ #1 \right\}}

\renewcommand{\d}{\partial}
\renewcommand{\half}{\frac{1}{2}}
\newcommand{\rcite}[1]{Ref. \cite{#1}}
\newcommand{\eqnref}[1]{Eq.~(\ref{#1})}
\newcommand{\secref}[1]{Section~\ref{#1}}
\newcommand{\appref}[1]{Appendix~\ref{#1}}
\newcommand{\figref}[1]{Fig.~\ref{#1}}
\newcommand{\sfigref}[2]{Fig.~\hyperref[#1]{\ref{#1}#2}}

\DeclareMathOperator{\Tr}{Tr}  

\renewcommand{\aa}{\boldsymbol{a}}
\newcommand{\bb}{\boldsymbol{b}}

\newcommand{\rr}{\boldsymbol{r}}
\newcommand{\kk}{\boldsymbol{k}}
\newcommand{\qq}{\boldsymbol{q}}

\newcommand{\hx}{\hat{x}}
\newcommand{\hy}{\hat{y}}

\newcommand{\hr}{\hat{\boldsymbol{r}}}
\newcommand{\hR}{\hat{\boldsymbol{R}}}
\newcommand{\hf}{\hat{f}}
\newcommand{\hg}{\hat{g}}
\newcommand{\hU}{\hat{U}}
\newcommand{\hu}{\hat{u}}
\newcommand{\hn}{\hat{n}}
\newcommand{\hphi}{\hat{\varphi}}

\newcommand{\PLLL}{P_{\text{LLL}}}

\renewcommand{\d}{\partial}

\newcommand{\signpost}[1]{{#1}}

\newcommand{\Caltech}{Department of Physics, California Institute of Technology, Pasadena, CA 91125}

\begin{document}

\preprint{APS/123-QED}

\title{Non-commutative effective field theory of the lowest Landau level superfluid}

\author{Nandagopal Manoj}\email{nmanoj@caltech.edu}\affiliation{\Caltech}

\date{\today}

\begin{abstract}
A 2+1D superfluid in a rapidly rotating trap forms an array of vortices, with collective excitations called Tkachenko modes. Reference~\onlinecite{Du2024a} argued from an effective field theory viewpoint that these excitations are described by a field theory living on a non-commutative space. We elucidate the microscopic origin of these non-commutative fields, and present a novel derivation of the effective field theory for this superfluid using a lowest Landau level projected coherent state path integral approach. Besides conceptual clarity, this approach makes quantitative predictions about the long-wavelength, low-energy behavior in terms of the microscopic parameters of the short-range interacting lowest Landau level superfluid — relevant to trapped Bose-Einstein condensate experiments.
\end{abstract}

\maketitle


\section{Introduction}
Following the discovery of the fractional quantum Hall effect~\cite{ExptTsuiStormerGossard1982}, it has been appreciated that two-dimensional quantum gases in strong magnetic fields produce exotic phases of matter originating from strong interactions engendered by quenched kinetic energy. Since then, a diverse set of approaches describing these states have emerged, including trial wavefunctions~\cite{Laughlin1983}, Haldane pseudopotentials~\cite{Haldane1983}, flux attachment~\cite{Wilczek1982,Jain1989}, Chern-Simons theory~\cite{LopezFradkin1991} for the gapped fractional quantum Hall states, and Halperin-Lee-Read (HLR) theory~\cite{HLR1993,Son2015} for the metallic phase of the half-filled Landau level. Field theoretic descriptions of these states such as HLR and Chern-Simons theory are notably not manifestly restricted to the lowest Landau level (LLL), even though the higher Landau levels are gapped out and, for the subset of these states that we are interested in, the relevant physics happens completely within the LLL as confirmed by numerical simulations~\cite{BookPrange2012Hall}. A natural question to ask is whether we can write down field theories in such a way that the LLL projection is manifest.

Non-commutative fields have been claimed to be useful to describe physics manifestly in the LLL since the late 1990s~\cite{Read1998,Polychronakos2001,SimeonHellerman2001,Susskind2001, Karabali2005, KarabaliNair2006, Pasquier2007} (for a review see Ref.~\cite{DouglasNekrasov2001}). Recent advances have exploited this technology to describe the $\nu=1$ bosonic composite Fermi liquid state~\cite{DongSenthil2020,DongSenthil2022,Goldman2022}, which uses an exact Pasquier-Haldane map~\cite{PasquierHaldane1998} to develop the non-commutative field theory. It has also been used to describe traditional Fermi liquids, where phase space is treated as a non-commutative space~\cite{Padayasi2025} . Another work~\cite{Du2024a} studied the LLL superfluid using boson-vortex duality~\cite{DasguptaHalperin1981} and argues that non-commutative field theories are the right language that describes the fluctuations above the mean-field vortex configuration. Using the non-commutative field theory, one can derive the correct scaling behavior of the dispersion relation and the decay rate of the low-energy excitations of this system, called Tkachenko modes~\cite{Tkachenko1966a,Tkachenko1966b,Tkachenko1969,Sonin2014,Sonin1976,Volovik1979,Baym2003}. The low energy theory is described by a quantum Lifshitz model (a similar theory has been used to describe vortex lattices in layered superconductors~\cite{BalentsRadzihovsky1996}). Although the mean-field vortex lattice state breaks both translation symmetry and $U(1)$ charge conservation symmetry spontaneously, this state only has one branch of low energy excitations, which disperses quadratically as $\omega \sim k^2$~\cite{WatanabeMurayama2013}.

This article solves the problem of how to arrive at a manifestly LLL-projected effective field theory (EFT) for the superfluid phase of bosons in the LLL, which can be experimentally realized in systems of Bose-Einstein condensates (BECs) in a rapidly rotating harmonic trap~\cite{BookCooperFQHBoson,ExptFletcherScience2021,ExptMukherjeeNature2022}. The problem of a rapidly rotating superfluid with a fine-tuned harmonic trap can be mapped at low energies to a LLL superfluid with an effective cyclotron frequency twice the trap rotation frequency. We refer to earlier works for readers interested in the explicit mapping~\cite{SinhaShlyapnikov2005}. In this work, we abstract away to the problem of bosons restricted to the LLL interacting via a contact repulsive interaction and derive an effective field theory for the LLL superfluid phase.  

Starting from a generic LLL-projected Hamiltonian for bosons describing a superfluid, we derive the non-commutative field theory of Ref.~\onlinecite{Du2024a} using physically motivated assumptions. We derive this effective description by developing a manifestly LLL-projected coherent state path integral. This approach both allows us to justify the non-commutative effective field theory for the LLL superfluid, and helps us connect these abstract, effective degrees of freedom to the microscopic contituents. In particular, for systems of BECs in rotating traps, which are composed of weakly interacting bosons, one can calculate the coefficients of the non-commutative Lagrangian in terms of the microscopic parameters and get quantitative predictions for the dispersion relation and decay rate of Tkachenko modes. We calculate the dispersion relation in this paper and leave the calculation of decay rate to future work.

The outline of the paper goes as follows: in \secref{sec:2}, we introduce the single-particle physics of the lowest Landau level and develop the machinery which we later use to define coherent states in \secref{sec:3}. In the process, we introduce the mathematics of non-commutative (NC) functions and refer the reader to \appref{app:Tools} for a more thorough discussion. In \secref{sec:3}, we develop the coherent state path integral restricted to the LLL that captures the effective degrees of freedom of the LLL superfluid, and in \secref{sec:4} effectively derive the non-commutative field theoretic action described in Ref.~\onlinecite{Du2024a}, using some symmetry arguments and dimensional analysis to arrive at the form of the most relevant interactions.

\section{Single particle physics in the LLL}
\label{sec:2}
We want to develop a field theoretic description of the LLL superfluid phase described by a manifestly LLL-restricted Hamiltonian of the form
\begin{widetext}
\begin{equation}
    H = \PLLL\int d^2\rr \term{-\mu b^\dagger(\rr)b(\rr) + \frac{g}{2}b^\dagger(\rr)b^\dagger(\rr)b(\rr)b(\rr)} \PLLL, \quad g>0, 
    \label{eqn:Ham1}
\end{equation}
\end{widetext}
where $\PLLL$ is a projection operator to the lowest Level and $b(\rr)$ is a local boson annihilation operator. We expect this type of interaction to quantitatively agree with the low-energy physics of rapidly rotating BECs.

\signpost{To do so, we begin by outlining the structure of single particle wavefunctions in the LLL. We describe the relevant single particle states for the LLL superfluid and explain how non-commutative (NC) functions arise as a natural language to identify these states. This will be useful to later define coherent states and develop a path integral for a many-body bosonic problem restricted to the LLL. }

To be concrete, we will work in Landau gauge $\boldsymbol{A}=(-By,0), B = l_B^{-2}$ but none of our results will specifically require this choice of gauge. In this gauge, a generic single particle wavefunction is given by 
\begin{equation}
    \psi(x,y) = f(z) e^{-y^2/2l_B^2}, \quad z = x+iy
\end{equation}
where $f(z)$ is a holomorphic function. Recall that the usual strip-like wavefunctions localized at $y=y_0$ in the Landau gauge are recovered using $f(z) = e^{-i y_0 (z - i y_0/2) / l_B^2}$.


A useful set of operators that act within the LLL are the projected position operators, also known as the guiding center operators $\hr = (\hx,\hy)$ that generate the full algebra of LLL operators, which in this gauge choice is defined as
\begin{align}
    \hx = x - i l_B^2 \frac{\partial}{\partial y}, \quad  \hy = i l_B^2 \frac{\d}{\d x},
\end{align}
satisfying $\comm{\hx,\hy}= -il_B^2$. In the rest of this section, we reserve the hat notation for the operators restricted to the LLL. The guiding center operators play a dual role in the LLL, indicated by their nontrivial commutation relation. $\bra{\psi}\hx\ket{\psi}$ measures the center of mass position of the LLL wavefunction $\ket{\psi}$ in the $x$ direction, like the usual position operator. On the other hand, its exponential $e^{iy_0 \hat{x} / l_B^2}$ acts as the magnetic translation operator in the $y$ direction. That is, $\hx$ also plays the role of the momentum in the $y$ direction. There is a similar picture for the $\hy$ operator, where everything is rotated by $90^\circ$. A general ``plane-wave'' function $e^{i\kk\cdot \hr}$ implements a magnetic translation of the wavefunction by $l_B^2 (\wedge \kk) \equiv (l_B^2 k_y, -l_B^2 k_x)$.

Due to the nontrivial commutation relations of the guiding center operators, these magnetic translation operators satisfy a closed algebra known as the Girvin-Macdonald-Platzman (GMP) algebra~\cite{GMP1986}. This can be derived using the Baker-Campbell-Hausdorff formula and is given by 
\begin{align}
    e^{i \kk \cdot \hr} e^{i \kk' \cdot \hr} &= e^{i l_B^2 \kk \wedge \kk'/2} e^{i(\kk+\kk')\cdot\hr} \\
    \comm{e^{i \kk \cdot \hr}, e^{i \kk' \cdot \hr}} &= 2 i \sin \term{l_B^2 \kk \wedge \kk'/2} e^{i(\kk+\kk')\cdot\hr}
\end{align}
where the wedge product is defined as $\kk\wedge \kk' = k_x k_y' - k_y k_x'$. Notably, the translation operators do not commute in the LLL. This algebra plays a crucial role in developing the non-commutative field theoretic description of the LLL.

\begin{figure}
  \centering
  \includegraphics[width=0.48\textwidth]{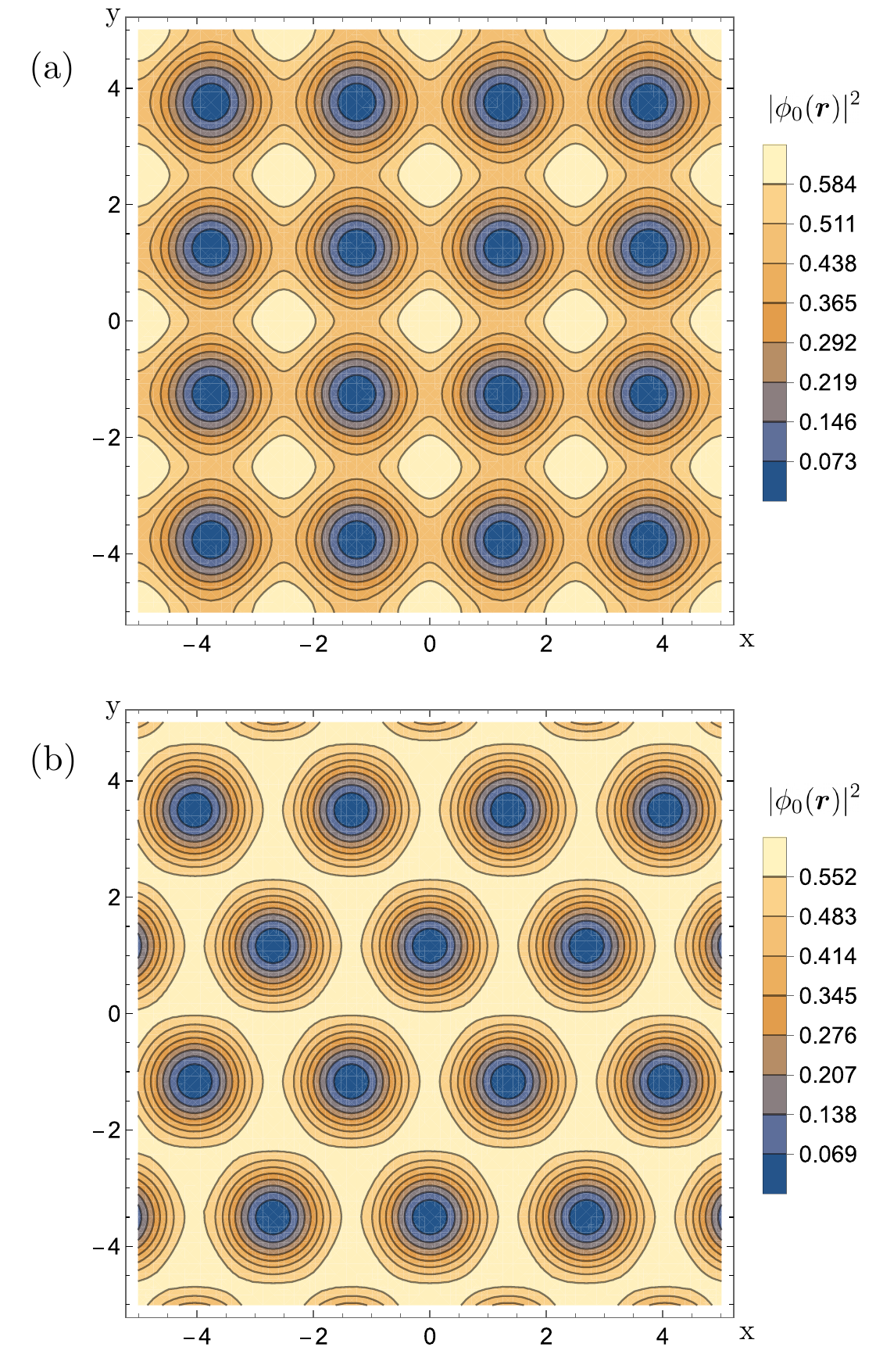}
  \caption{Plot of $\abs{\Phi_0(\rr)}^2$ for (a) square lattice and (b) triangular lattice (in units where $l_B = 1$). The blue cores are the vortices.}
  \label{fig:1}
\end{figure}

Although general translation operators do not commute, two translation operators commute if the parallelogram formed by the real-space translation vectors $l_B^2 (\wedge \kk)$ encloses an integer quantum of magnetic flux. Let us choose a pair of translation operators such that this integer is one, i.e.,
\begin{align}
    \hat{T}_1 = e^{i \kk_1 \cdot \hr}, & \quad
    \hat{T}_2 = e^{i \kk_2 \cdot \hr}, \nonumber \\
    \text{s.t. }\kk_1\wedge \kk_2 &= 2 \pi l_B^{-2}.
    \label{eqn:Lattice_Quantization}
\end{align}
Simultaneously diagonalizing these operators allows us to find magnetic lattice translation invariant wavefunctions where the zeros of the holomorphic function $f(z)$ are arranged in a Bravais lattice defined by the translation vectors $\aa_i = l_B^2 (\wedge \kk_i)$. We can also interpret the zeros of the wavefunctions as vortices, since the wavefunction winds by $2\pi$ as one traverses around it. So, we will call such wavefunctions vortex lattices. The specific form of the holomorphic functions is irrelevant for the discussion -- they can be constructed using elliptic theta functions or Weierstrass sigma functions~\cite{HaldaneRezayi1985}. One useful and intuitive construction of these functions is using the strip-like wavefunctions. For the case of square lattice $\aa_1 = (\sqrt{2\pi} l_B,0),\aa_2 = (0, \sqrt{2\pi} l_B)$, the wavefunction $\ket{\Phi_0}$ with eigenvalue one for both $\hat{T}_1$ and $\hat{T}_2$ can be written as 
\begin{equation}
   \Phi_0(\rr) \propto  \sum_{y_0 = \sqrt{2\pi}l_B\mathbb{Z}} e^{-i y_0 (z - i y_0/2) /l_B^2} e^{-y^2/2l_B^2},
   \label{eqn:Phi_square}
\end{equation}
a uniformly spaced uniform superposition of the strip-like wavefunctions. \changes{Similarly, the triangular lattice wavefunction with translation vectors $\aa_1 = (\sqrt{\sqrt{3}\pi} l_B,\sqrt{\pi/\sqrt{3}}l_B),\aa_2 = (\sqrt{\sqrt{3}\pi} l_B, - \sqrt{\pi/\sqrt{3}}l_B)$ can be written as 
\begin{equation}
   \Phi_0(\rr) \propto  \sum_{y_0 = \sqrt{\sqrt{3}\pi}l_B\mathbb{Z}} (-1)^{\frac{j(j+1)}{2}} e^{-i y_0 (z - i y_0/2) /l_B^2} e^{-y^2/2l_B^2}.
   \label{eqn:Phi_triangular}
\end{equation}
Both these wavefunctions have been plotted in \figref{fig:1}.}
The other eigenfunctions of the translation operators can be obtained by applying the magnetic translation operator on $\ket{\Phi_0}$. We can use the GMP algebra to show that $\ket{\Phi_{\qq}} = e^{i \qq \cdot \hr} \ket{\Phi_0}$ has eigenvalue $e^{i \kk_i\wedge \qq l_B^2} = e^{-i \aa_i \cdot  \qq}$ under $\hat{T}_i$. The $\ket{\Phi_{\qq}}$ form a complete orthonormal basis for the LLL wavefunctions. 

It is a useful perspective to think of the LLL as a regulator for a field theory analogous to a lattice, with Bravais vectors $\aa_i$. In this sense, $\ket{\Phi_{\qq}}$ are analogous to Bloch wavefunctions living on a two-torus Brillouin zone and $l_B^{-1}$ serves as a UV cutoff while preserving continuous translation and rotation symmetry. Crucially, the band containing these single particle states is topological and has Chern number one, which means that we cannot define the single particle wavefunctions $\Phi_{\qq}(\rr)$ in a continuous single-valued form over the Brillouin zone. 

\changes{Note that we are free to choose any Bravais lattice that satisfies the quantization condition (\ref{eqn:Lattice_Quantization}),} but for typical applications for the many-body problem it is energetically favorable to work with the triangular lattice. \changes{The formalism developed in this article is agnostic to the choice of lattice, and in the end we will briefly comment about the case of the square lattice}. 

A brief aside on the properties of operators in the LLL, denoted $\hf(\hr)$. We will call this a non-commutative function or a non-commutative field, interchangeably depending on the context \changes{(we refer the reader to Ref.~\cite{DouglasNekrasov2001} for an extensive review, and \appref{app:Tools} for a brief overview of the calculus of non-commutative functions)}. This is an arbitrary operator acting within the LLL as the guiding-center operators generate the full algebra of the LLL. Since it is a linear operator, it is itself also part of a vector space. \changes{In our analysis,} we assume that the functions are sufficiently well-behaved such that a suitable choice of basis for this vector space are plane wave functions $e^{i \kk \cdot \hr}$. This forms a complete basis for this vector space, i.e., one can write any NC function $\hf(\hr) = \sum_{\kk} f_{\kk} e^{i\kk \cdot \hr}, f_{\kk}\in \mathbb{C}$. We will call $f_{\kk}$ the Fourier transform of $\hf$. Note that this Fourier transform takes a NC function to a regular function. 

We start with $\ket{\Phi_0}$ as our reference state. We define 
\begin{equation}
    \ket{\hf} = \hf(\hr)\ket{\Phi_0}
\end{equation}
where $\hf$ is an arbitrary function of the guiding center operators. Since the guiding center operators generate the full algebra of the LLL \changes{(similar to the position and momentum operators in a one-dimensional quantum mechanics problem)}, we claim\changes{\footnote{\changes{We do not rigorously prove this statement as that requires us to go into details about the space of allowed functions, which is beyond the scope of this work.}}} that for any $\ket{\psi} \in \mathcal{H}_{\text{LLL}}$, there exists an $\hf$ such that $\ket{\psi} = \hf(\hr)\ket{\Phi_0}$. Notably, this $\hf$ is not unique\footnote{This is also straightforwardly seen by noting that the vector space of $\hf$ has dimension $\text{dim}(\text{LLL})^2$.}, any operator that leaves $\ket{\Phi_0}$ invariant can be right-multiplied to $\hf$ to get a different function giving rise to the same LLL state. An example of such a NC function is the translation operator $\hat{T}_1 = e^{i \kk_1 \cdot \hr}$ satisfying $\hat{T}_1\ket{\Phi_0} = \ket{\Phi_0}$. To get rid of this ambiguity, we make a physical approximation -- we set a hard momentum cutoff $\Lambda \ll l_B^{-1}$ and only consider functions whose Fourier transforms $f_{\kk}$ are non-zero only if $\abs{\kk} < \Lambda$. Recalling that plane-wave NC functions implement translations of the vortex lattice proportional to its momentum, this restriction means that we only consider weak, long wavelength deformations of the vortex lattice. Using the analogy to lattice regularization, this is completely natural. When we describe a system using a field theory, we implicitly assume that the fields fluctuate at momenta much smaller than the UV cutoff -- we are making the analogous approximation here.

Upon making this approximation, the previously mentioned ambiguity disappears. \changes{We note that there are no long-wavelength (low momentum) NC functions that leave $\ket{\Phi_0}$ invariant except the trivial $\hf(\hr) = 1$, as 
\begin{equation}
    \ket{\hf} = \sum_{\kk} f_{\kk} e^{i \kk\cdot \hr} \ket{\Phi_0} = \sum_{\kk} f_{\kk} \ket{\Phi_{\kk}},
\end{equation}
and $\sum_{\abs{\kk}<\Lambda} f_{\kk} \ket{\Phi_{\kk}} = \ket{\Phi_0}$ iff $f_{\kk} = \delta_{\kk,0}$ due to orthonormality of the LLL Bloch wavefunctions.} This restriction also saves us from worrying about the phase ambiguity of the basis states as the LLL is a Chern band with Chern number one, but since $f_{\kk}$ is zero close to the edges of the first Brillouin zone and beyond, we do not have to worry about the phase winding of the LLL Bloch wavefunctions and can unambiguously define what we mean by every state $\ket{\Phi_{\kk}}$. 

There is a crucial downside to working in the long-wavelength approximation, which is forced on us. This is simply that we should start from the ``correct'' reference state. That is, if we start from a square vortex lattice but the true ground state is close to a triangular lattice, our long-wavelength deformations cannot reach that state and it remains inaccessible. Instead, we will only see the physics of small deformations around the square lattice. The triangular lattice is in a different vacuum superselection sector. 

Another feature of the NC functions is that if one looks at the appropriately normalized inner product
\begin{equation}
    \braket{\Phi_0}{\rr}\bra{\rr} e^{i \kk \cdot\hr} \ket{\Phi_0} = \Phi_0^\ast(\rr) e^{i \kk \cdot\hr} (\Phi_0(\rr)),
\end{equation}
it looks like it is equal to $e^{i \kk \cdot \rr}$ times some short-wavelength modulation due to the zeros of the wavefunction \changes{(cf.~\figref{fig:1})}, where $\kk$ is a small momentum (proved in \appref{app:CGIP}). In principle, one can coarse-grain away the short wavelength modulations and this coarse grained function really acts like a regular plane-wave acting on a uniform wavefunction. A simple corrollary for a long wavelength NC function is that 
\begin{equation}
    \braket{\Phi_0}{\rr}\bra{\rr} \hf(\hr) \ket{\Phi_0} \approx \sum_{\kk} f_{\kk} e^{i \kk \cdot \rr} 
\end{equation}
due to linearlity (the $\approx$ denotes coarse-graining and ignoring higher-order terms in $\abs{\kk}^2l_B^2$). This motivates a mathematical map between NC functions and regular functions, known as the Weyl map~\changes{\cite{DouglasNekrasov2001}}. The Weyl map is a linear, invertible map given by
\begin{equation}
    W\comm{e^{i \kk \cdot \hr}} = e^{i \kk\cdot \rr}.
\end{equation}
Then, we can write
\begin{equation}
    \braket{\Phi_0}{\rr}\bra{\rr} \hf(\hr) \ket{\Phi_0} \approx W\comm{\hf}(\rr).
\end{equation}
We will sometimes use the natural notation $W[\hf] = f$. The commutative function $f$ is also called the Weyl symbol of $\hf$.

We can calculate the inner product
\begin{align}
    \braket{\hf}{\hg} &= \sum_{\kk,\kk'} f^\ast_{\kk} g_{\kk'} \bra{\Phi_0}e^{-i \kk \cdot \hr} e^{i \kk' \cdot\hr} \ket{\Phi_0} \\
    &=\sum_{\kk} f^\ast_{\kk} g_{\kk} \\
    &= \Tr\comm{\hf^\dagger \hg} \\
    &= \int d^2 \rr f^\ast(\rr)g(\rr)
\end{align}
where the last integral is a convenient expression using the Weyl symbols for the NC function. For the trace notation, refer to \appref{app:Tools}. In the second step, we also used the fact that $\hf$ and $\hg$ have support only at low momenta, so the only way that inner product could be non-zero is if $\kk' - \kk = 0$.

Finally, we describe how magnetic translation symmetry acts on these wavefunctions. This is the most important feature of this formalism and it makes NC fields useful in the first place. The form of the Weyl map tells us that there is no fundamental difference between an NC field and a regular field -- there is an invertible map that takes a long-wavelength NC field to a long-wavelength commutative field and we can use either function to uniquely describe a state. The advantage of working with NC fields is that translation symmetry is manifest in this formalism, which is crucial when writing down effective field theories. In the earlier discussion, we noted that magnetic translation symmetry is implemented in the LLL using the plane-wave functions -- the operator $e^{i\kk\cdot \hr}$ shift the wavefunction by $\aa = l_B^2(\wedge\kk)$. This means that 
\begin{align}
    T_{\aa}\ket{\hf} &= e^{i\kk\cdot \hr}\ket{\hf} \\
    &= (e^{i\kk\cdot \hr}\hf)\ket{\Phi_0} \\
    &= \ket{e^{i\kk\cdot \hr}\hf},
\end{align}
that is the state labelled by $\hf(\hr)$ maps to the state labelled by $(e^{i\kk\cdot \hr}\hf(\hr))$ upon translation. Note that this is not equal to $\hf(\rr-\aa)$, contrary to the naive expectation. On the other hand, the Weyl symbol transforms as
\begin{align}
    f(\rr) &\mapsto W\comm{e^{i\kk\cdot \hr}\hf(\hr)} \\
    &= \sum_{\kk'} f_{\kk'} W\comm{e^{i\kk\cdot \hr}e^{i\kk'\cdot \hr}} \\
    &= \sum_{\kk'} f_{\kk'} e^{i \frac{l_B^2}{2} \kk\wedge \kk'} e^{i(\kk+\kk')\cdot \rr} \\
    &= e^{i \kk \cdot \rr} f\term{\rr - \frac{l_B^2}{2}(\wedge\kk)} \\
    &= e^{i \kk \cdot \rr} f\term{\rr - \frac{\aa}{2}}.
\end{align}
It is not clear from the transformation of the Weyl symbol that this corresponds to a physical translation by $\aa$. If we want to write down a manifestly translation symmetric effective field theory describing the LLL superfluid, it is better to work with variables that transform nicely under these symmetries, namely the NC functions $\hf$.

\section{Many body bosonic coherent states in the LLL}
\label{sec:3}
\signpost{In the previous section, we developed some machinery to describe single particle states, using NC functions. Next, we use these single particle states to construct bosonic coherent states by forming coherent superpositions of the same single-particle state at different occupation number. Our goal will be to use this manifestly LLL coherent state basis to construct a path integral for the interacting boson problem, cf.~\eqnref{eqn:Ham1}.} We define the bosonic coherent state using the local (not LLL-projected) boson creation operator $b^\dagger(\rr)$
\begin{equation}
    \Ket{\hf} \equiv e^{\int d^2 \rr \braket{\rr}{\hf} b^\dagger(\rr)}\Ket{0}
\end{equation}
where we use the $\Ket{}$ notation for Fock (second-quantized) states and $\Ket{0}$ is the vacuum state with no particles. We claim that 
\begin{enumerate}
    \item These states live purely in the LLL. This is because $b^\dagger_{\hf} \equiv \int d^2 \rr \braket{\rr}{\hf} b^\dagger(\rr)$ adds a boson into the LLL state $\hf\ket{\Phi_0}$. Since $\Ket{\hf}$ is a coherent sum of many particles in this same single particle orbital, this state is manifestly in the LLL. 
    \item These states form an (over)complete basis for the many body LLL states. We can write 
    \begin{equation}
        b^\dagger_{\hf} = \left. \frac{d}{d\alpha}e^{\int d^2 \rr \braket{\rr}{(\alpha\hf)} b^\dagger(\rr)} \right\vert_{\alpha = 0},
    \end{equation}
    and since $\ket{\hf}$ spans the single particle LLL Hilbert space, we can use this to construct any Fock state in the occupation number basis. Since the different $b^\dagger(\rr)$ commute, the products of $b^\dagger_{\hf}$ constructed this way remain in the ``sum of exponentials'' form that we need, and therefore any Fock state can be written as a superposition of many $\Ket{\hf}$.
\end{enumerate}

With these properties, we can develop the standard machinery for a coherent state path integral but now completely restricted to the LLL. These coherent states satisfy 
\begin{equation}
    b(\rr)\Ket{\hf} = \braket{\rr}{\hf}\Ket{\hf} \implies b_{\hg} \Ket{\hf} = \braket{\hg}{\hf}\Ket{\hf}
\end{equation}
and have inner product 
\begin{align}
    \Braket{\hf}{\hg} &= e^{\braket{\hf}{\hg}} \\
    &= e^{\Tr\comm{\hf^\dagger \hg}} \\
    &= e^{\int d^2\rr f^\ast(\rr)g(\rr)}
\end{align}
The resolution of identity is given by 
\begin{align}
    \mathbb{1}_\mathcal{F} = \int D\hf e^{-\braket{\hf}{\hf}}\Ket{\hf}\Bra{\hf}
\end{align}
which can be proved by verifying that $\mathbb{1}_\mathcal{F}\Ket{\hg} = \Ket{\hg}$ for any long-wavelength $\Ket{\hg}$. The trace of an second-quantized operator restricted to the LLL is given by 
\begin{align}
    \Tr[O] = \int D\hf e^{-\braket{\hf}{\hf}}\Bra{\hf} O \Ket{\hf}.
\end{align}
Functional integrals over NC fields have been defined in \appref{app:Tools}. We recommend that any reader unfamiliar with NC field theory works through that appendix before continuing to the rest of the article. It is important to emphasize here that this identity and trace operations act only within the subspace of states spanned by long wavelength NC fields $\Ket{\hf}$. This restriction is justified as the path integral is energetically restricted to this space at low energies in the physical problem. For more general LLL problems, one cannot blindly use this dictionary to construct path integrals -- the long-wavelength restriction must be justified. 

Now that we have the main ingredients, we can trotterize the Hamiltonian (\ref{eqn:Ham1}) to get a functional integral for the partition function, completely restricted to the LLL (we use $\Tr_\mathcal{F}$ to denote trace over Fock space, not to be confused with the trace over non-commutative space). Recall the Hamiltonian
\begin{widetext}
\begin{equation}
    H = \PLLL\int d^2\rr \term{-\mu b^\dagger(\rr)b(\rr) + \frac{g}{2}b^\dagger(\rr)b^\dagger(\rr)b(\rr)b(\rr)} \PLLL, \quad g>0.
\end{equation}
Then, the partition function is given by
\begin{align}
    Z &= \Tr_\mathcal{F}\comm{e^{-\beta H}} \\
    &= \Tr_\mathcal{F}\comm{e^{-\Delta\tau H} \mathbb{1}_\mathcal{F}e^{-\Delta\tau H} \mathbb{1}_\mathcal{F}\dots e^{-\Delta\tau H} \mathbb{1}_\mathcal{F}} \qquad \text{($N$ times)} \\
    &= \int D\hf_0 D\hf_1 \dots D\hf_{N-1} e^{-\sum_{j=0}^{N-1} \braket{\hf_j}{\hf_j}} \Bra{\hf_0}e^{-\Delta\tau H} \Ket{\hf_{N-1}} \Bra{\hf_{N-1}}e^{-\Delta\tau H} \Ket{\hf_{N-2}} \dots \Bra{\hf_1}e^{-\Delta\tau H} \Ket{\hf_{0}} \\
    &\xrightarrow{N\to \infty} \int D\hf \exp\comm{-\int_0^\beta d\tau  \comm{\bra{\hf_\tau} \partial_\tau \ket{ \hf_\tau} - \int d^2\rr \Bra{\hf_\tau} H \Ket{\hf_\tau}}} \\
    &= \int D\hf \exp\comm{-\int_0^\beta d\tau  \comm{\bra{\hf_\tau} \partial_\tau \ket{ \hf_\tau} - \int d^2\rr \term{-\mu \braket{\hf_\tau}{\rr}\braket{\rr}{\hf_\tau} + \frac{g}{2} \braket{\hf_\tau}{\rr}\braket{\hf_\tau}{\rr}\braket{\rr}{\hf_\tau}\braket{\rr}{\hf_\tau}}}} \label{eqn:microscopic_action}\\
    &= \int D\hf \exp\comm{-\int_0^\beta d\tau  \Tr\comm{\partial_\tau {\hf}  \hf^\dagger -\mu {\hf}  \hf^\dagger + \frac{g^\ast}{2} {\hf}  \hf^\dagger{\hf}  \hf^\dagger + \dots }}, \qquad \hf \equiv \hf(\hr,\tau). \label{eqn:effective_action1}
\end{align}
\end{widetext}
Note that the projection to the LLL comes for free as the coherent states are LLL-restricted, and that the NC field $\hf$ describes a boson. The last step is paramount. The time derivative and chemical potential terms match exactly when we rewrite it in terms of the trace over non-commutative space, but the quartic term cannot be exactly mapped in that way. From the long-wavelength behavior of the wavefunction products (\appref{app:CGIP} and discussions earlier), we know that it is approximated by the leading order term written above, where $g^\ast ( \approx 1.160 g$ for the triangular vortex lattice\changes{~\cite{YangMacdonald2004}, details in \appref{app:C}}) is a renormalized interaction vertex, which can be computed by numerically estimating the interaction energy of the reference state $\sim \int d^2 \rr \abs{\Phi_0(\rr)}^4$. Since this is not an exact mapping, there are corrections that have to be quartic in the field operators because of the scaling with density, and they come in the form of higher derivative terms in the NC field theory language. These corrections depend on the exact vortex lattice reference state $\ket{\Phi_0}$ that we start with, and we will not try to calculate the coefficients from microscopics. Rather, we use symmetry and dimensional analysis to constrain the form of the correction terms. 

The Lagrangian $\mathcal{L} = \partial_\tau {\hf}  \hf^\dagger +\mu {\hf}  \hf^\dagger + \frac{g^\ast}{2} {\hf}  \hf^\dagger{\hf}  \hf^\dagger + \dots $ is now a NC function. It does not have a spatial derivative term, and any unitary, time-independent $\hf$ has the same action. This situation is clearly not physical. Another way to see this is that this action is invariant under any time-independent symmetry transformation $\hf \mapsto \hu(\hr) \hf$. 
One might think of adding the manifestly symmetry covariant term $\partial_i(\hf \hf^\dagger)\partial_j(\hf \hf^\dagger)$ or higher derivatives acting on the uncharged operators. This additional term is allowed, but does not do anything substantial as any unitary $\hf$ does not cost energy under this term. In the effective theory of the superfluid that we write down later, this term will not generate a dispersion for the Tkachenko mode.

We find that there is no derivative four-field term that both respect the spatial symmetries and gives an energy cost to unitary $\hf$. There is a similar issue if one tries to write down a local field theory in terms of the boson field operators for a dipole conserving Bose-Hubbard model~\cite{Lake2022} which captures the effects of dipole hopping. Another approach is to start from a non-local\footnote{Non-local here just means that the terms in the Lagrangian are not simply a function of a field and its derivatives at the same spatial point. The theory is still physically local, it has exponential decay of interactions.} hopping term in the Lagrangian 
\begin{equation}
    \int_{{\aa},{\bb}}g({\aa},{\bb})\hf(\hr) \hf^\dagger(\hr+ \aa) \hf(\hr + \aa + \bb) \hf^\dagger(\hr + \bb ), 
\end{equation}
where $g(\aa,\bb) = g \delta(\aa\cdot \bb )\mathcal{O}(e^{-(\aa^2 + \bb^2)l_B^{-2}})$ is a decaying function reflecting the locality of the hopping process. This form of the hopping term is constrained by magnetic translation and rotation symmetry, which implies dipole moment $\int \rho \hat{r}_i$ and trace of quadrupole moment $\int \rho \hat{r}_i \hat{r}_i/2$ conservation respectively, as these are the generators of those symmetries~\cite{Spodyneiko2023}. Then, we use Hubbard-Stratonovich fields (\appref{app:Tools}) to resolve the interaction terms into quadratic pieces. Because of the symmetries of the problem, we have to introduce two sets of Hubbard-Stratonovich fields, one for the trace of quadrupole moment and a vector field for the dipole moment. Using this, we can systematically form condensates of each of these fields. The mean-field phases naively correspond to
\begin{enumerate}
    \item \textit{Trivial insulator}: a phase with all the symmetries unbroken and the ground state is smoothly connected to the bosonic vacuum state. 
    \item \textit{Hexatic phase}: a phase with spontaneously broken rotation symmetry and a linearly dispersing Nambu-Goldstone boson (NGB). 
    \item \textit{Crystalline phase}: a phase with spontaneously broken rotation and translation symmetry, with a quadratically dispersing NGB. The naive expectation of two linearly dispersing NGB is subverted due to the symmetry allowed $\epsilon_{ij} d_i \partial_\tau d_j$ term, where $e^{id_j}$ is the two-component dipole/translation order parameter. This is similar to the NGB physics of a Heisenberg ferromagnet, and is a type-B NGB~\cite{WatanabeMurayama2013,Beekman2019}.
    \item \textit{LLL superfluid phase}: a phase with spontaneously broken rotations, translations and $U(1)$ charge conservation symmetry. This corresponds to a superfluid vortex lattice and has a single quadratically dispersing NGB (related to the Tkachenko mode). Although it is quadratically dispersing, its physics is qualitatively different from the NGB of the crystalline phase as here it is a type-A NGB.
\end{enumerate}
{Note that the crystalline phase here does not refer to LLL Wigner crystal states, which are expected to exist at low boson densities and is beyond the scope of this work. We specifically refer to the phases obtained using the Hubbard-Stratonovich field treatment in this section, which in this case will be a precursor to the LLL superfluid phase with the same pattern of broken translation symmetry, i.~e. density variations at the scale of $l_B$. This is in contrast with the Wigner crystal whose density varies at the scale of $n^{-1/2}$, where $n$ is the boson density.}

The naive mean-field theory predicts that there is a cascade of transitions between the trivial insulator and the LLL superfluid via spontaneous breaking of various higher moments of charge conservation symmetry. This naive treatment is wrong, and there is a simple reason why. Take the crystalline phase, for example. This phase can be thought of as a dipole condensate, that is, a phase where dipoles are proliferated starting from the parent trivial insulator. A necessary condition for this to be true is that the average charge density of this state is the same as the trivial insulator phase -- which has zero density. But we know that the bosonic vacuum is the only state with exactly zero density, and there are no negative density states\footnote{An identical issue arises around the zero filling Mott insulator lobe of the dipolar Bose-Hubbard model and is discussed in footnote 2 of \cite{Lake2022}.}. Using an analogous argument for the hexatic phase (which is a condensate of quadrupoles), we conclude that there are no non-trivial hexatic and crystalline phases in this problem, and we generically expect a first-order transition between the trivial insulator phase and the LLL superfluid phase. \changes{Refs.~\cite{NguyenMoroz2023,Du2024b} study these phases from the dual perspective, i.~e. in terms of melting of the vortex lattice. Under the identification of the vortex‐liquid with the trivial phase introduced above, their findings appear to contradict our results. Ref.~\cite{Du2024b} describes a series of transitions connecting the vortex solid (LLL superfluid) to the vortex liquid (trivial) phase, via the intermediate phases described above. This disagrees with our results as we argue that these intermediate phases do not exist, which becomes apparent when viewed from the original (non-dual) language. Ref.~\cite{NguyenMoroz2023} speculates about a continuous direct transition between the vortex solid and vortex liquid phase, but we argue that such a transition will generically be first order. This is because from our perspective, the vortex liquid phase is the same as the bosonic vacuum, and one cannot have a continuous transition between states with different boson density. The disagreement is resolved when we note that the ideas in Refs.~\cite{NguyenMoroz2023,Du2024b} become relevant when we observe transitions between bosonic quantum Hall states and LLL superfluids, where the quantum Hall state is a vortex liquid phase at non-zero boson density, distinct from the bosonic vacuum state we consider here. We leave a microscopic understanding to the effective field theory describing this transition to future work.}

\section{Non-commutative field theory of the LLL superfluid}
\label{sec:4}
\signpost{With the form of the effective non-commutative Lagrangian for the interacting bosons in the LLL, we are ready to reproduce the effective NC field theory action for the LLL superfluid described in \rcite{Du2024a} by integrating out the amplitude mode. Like discussed earlier, we will use some symmetry and dimensional analysis arguments to arrive at the term that provides a dispersion for the Tkachenko mode.} 

To integrate out the amplitude fluctuations, we perform a polar decomposition of our field operator $\hf = \hu \sqrt{\hn}$, and write $\hn = n_0 + \delta\hn$. Here, $\hu$ is a unitary matrix and $\hn$ is a positive semi-definite Hermitian matrix. Since only small $\delta\hn$ are energetically accessed in the path integral, we can integrate over $\delta \hn$ assuming that it is a real NC field, with no other constraints. Like the analogous procedure in regular superfluids, we ignore derivatives of $\delta \hn$. 
\begin{widetext}
\begin{align}
    Z &= \int D\hf \exp\comm{-\int d\tau  \Tr\comm{\partial_\tau {\hf}  \hf^\dagger - \mu {\hf}  \hf^\dagger + \frac{g^\ast}{2} {\hf}  \hf^\dagger{\hf}  \hf^\dagger +\dots }} \\
    \begin{split}
    &= \int D\delta\hn D\hu \exp\Biggl[ -\int d\tau  \Tr \biggl[ \term{\mu n_0 - \frac{g^\ast}{2} n_0^2} + \term{n_0 \partial_\tau \hu \hu^\dagger }  + \delta\hn\term{\hu^\dagger \partial_\tau \hu - \mu + g^\ast n_0 } + \frac{g^\ast}{2}\delta\hn^2 + \dots \biggr] \Biggr]. 
    \end{split}
\end{align}
\end{widetext}
We further ignore the $\delta\hn$ contributions coming from the $(\dots)$ terms. This action is minimized at mean-field theory when $n_0 = \mu / g^\ast$, so we make that replacement. Then, we integrate out $\delta\hn$ using the formula for Gaussian integrals discussed in \appref{app:Tools}. Upon doing so, and ignoring total derivative terms, we get
\begin{widetext}
    \begin{align}
        \frac{Z}{Z_{\text{MF}}} &= \int D\hu \exp\comm{-\int d\tau \Tr\comm{\frac{\mu}{g^\ast} \partial_\tau \hu \hu^\dagger - \frac{1}{2g^\ast} \term{\hu^\dagger \partial_\tau \hu }^2  + \dots }}.
    \label{eqn:EFT1}
    \end{align}
\end{widetext}
Let us briefly talk about the path integral measure $D\hu$. One natural way to define this measure is to write $\hu = e^{i\hphi}$ and use the measure for real NC field $D\hphi$, and this is what we will do. When doing so, we ignore the effect of topological defects -- unitary transformation paths that are not smoothly connected to the identity operator. This is justified if we only care about small deformations away from the vortex lattice reference state. 

Focusing on the action, we note that the first term is invarant under time-reparametrization symmetry $\hu(\tau) \mapsto \hu(\alpha(\tau))$, where $\alpha$ is a differentiable map. This means that the first term is a geometric phase, since it only depends on the path we take and not how quickly we traverse the path. This is the non-commutative \emph{Berry Lagrangian} term, discussed previously in a commutative field theoretic description of a LLL superfluid~\cite{MorozSon2019}. It encodes the Aharanov-Bohm phase picked up by the superfluid bosons due to the magnetic field. One way to see that is by noting that $\hu$ is a unitary map in the LLL, which in the Heisenberg picture encodes an area preserving deformation (APD) of space~\cite{Du2022}, given by
\begin{equation}
    \hr \mapsto \hR = \hu \hr \hu^\dagger.
\end{equation}
Following \rcite{Du2024a}, this map $\hR(\hr)$ can be interpreted as the non-commutative Eulerian coordinate map~\cite{soper2008classical} of the deformation of the reference vortex lattice. The Aharonov-Bohm phase picked up by a closed path $\hu(t)$ will be given by (in the NC field theory language)
\begin{align}
    \int dt \Tr\comm{ \frac{n_0 l_B^{-2}}{2}\epsilon_{ij}\hat{R}_i \partial_t \hat{R}_j } &= \int d\tau \Tr\comm{ i n_0 \partial_t \hu \hu^\dagger}
\end{align}
which is the real-time version of the first term in the effective action (\ref{eqn:EFT1}). 

The second term describes the cost of temporal fluctuations of the Tkachenko mode, acting as an interaction-induced kinetic energy in the LLL.
To get the potential energy, we use the insight from elasticity theory and the coordinate map $\hR(\hr)$. Choosing the reference state $\ket{\Phi_0}$ that minimizes this potential energy (a triangular lattice), we know that the potential energy of a configuration $\hu$ due to the local interactions between bosons will be approximated by
\begin{equation}
    \half g^\ast n_0^2 C_{ijkl} \hu_{ij} \hu_{kl}.
\end{equation}
to the lowest order in derivatives. Here, $C_{ijkl}$ is a dimensionless elastic modulus tensor and $\hu_{ij}$ is the non-commutative strain tensor, given by
\begin{equation}
    \hu_{ij} = \half\term{\d_i \hu_j + \d_j \hu_i},
\end{equation}
(valid for small deformations) where $\hu_i(\hr) = \hat{r}_i - \hat{R}_i(\hr)$ is the displacement field, not to be confused with the unitary field $\hu$. The displacement field can be simplified as 
\begin{align}
    \hu_i(\hr) &= \hat{r}_i\hu \hu^\dagger - \hu \hat{r}_i \hu^\dagger \\
    &= il_B^2 \epsilon_{ij} \partial_j\hu \hu^\dagger
\end{align}
which gives the strain tensor
\begin{equation}
    \hu_{ij} = \frac{il_B^2 \epsilon_{jj'}}{2}\term{\partial_i(\partial_{j'}\hu\hu^\dagger)} + (i\leftrightarrow j )
\end{equation}
To get the free theory of the Tkachenko mode, we expand the Lagrangian to quadratic order in $\hphi$, where $\hu = e^{i\hphi}$. We will describe the action in the real-time language $\tau \to it$. Going forward, we use the symmetries of the elastic tensor $C_{ijkl} = C_{jikl} = C_{ijlk} = C_{klij}$ to simplify expressions. 
\begin{widetext}
    \begin{align}
        S[\hphi] &= {\int dt \Tr\comm{-i\frac{\mu}{g^\ast} \partial_t \hu \hu^\dagger + \frac{1}{2g^\ast} \term{-\hu^\dagger \partial_t \hu }^2 - \frac{g^\ast n_0^2 C_{ijkl} i^2 l_B^4 \epsilon_{jj'}\epsilon_{ll'} }{2}\partial_i(\d_{j'} \hu\hu^\dagger)\partial_k(\d_{l'}\hu\hu^\dagger) }} \\
        &= \int dt \Tr\comm{-i\frac{\mu}{g^\ast} \term{i\d_t \hphi + \frac{\d_t \hphi \hphi - \hphi\d_t\hphi}{2}} + \frac{1}{2g^\ast} \term{-i\d_t\hphi}^2 -\frac{g^\ast n_0^2 C_{ijkl} l_B^4 \epsilon_{jj'}\epsilon_{ll'} }{2}\partial_i\d_{j'} \hphi\partial_k \d_{l'}\hphi + \mathcal{O}(\hphi^3)}\\
        &= \int dt \Tr\comm{ -\frac{1}{2g^\ast}(\d_t\hphi)^2 - \frac{\mu^2 \lambda l_B^4}{2g^\ast}(D_{ij} \hphi)^2  }, \qquad D_{ij} = \d_i \d_j - \half \delta_{ij} \nabla^2. 
    \label{eqn:EFT2}
   \end{align}
\end{widetext}
$\lambda \approx 0.29 $ for the triangular lattice is computed by numerically estimating the interaction energy for the LLL coherent state $\Ket{\hu}$, with 
\begin{equation}
    \hu = \exp(i \varphi_0 \cos(\kk \cdot \hr)); \quad \abs{\kk}l_B, \varphi_0 \ll 1.
    \label{eqn:deformation}
\end{equation} 
That is, we compute the energy cost $\int d^2\rr g n_0^2 \abs{\Phi_{\hu}(\rr)}^4$ numerically and match it with the effective field theory to get the correct coefficients. \changes{The details are shown in \appref{app:C2}.} This quadratic low-energy action is valid as long as the system is short-range interacting, and at the thermodynamic and LLL limit. The Tkachenko field $\hphi$ describes both the coarse-grained phase fluctuations of the LLL superfluid, and (its gradient describes) phonon fluctuations of the vortex lattice~\cite{Du2024a}. The Berry Lagrangian term is a total derivative term at quadratic order, and does not affect the dispersion relation of the Tkachenko mode, but this term is crucial for the transport physics of the LLL superfluid~\cite{Du2024b}. We have reproduced the quadratic EFT described in Ref.~\onlinecite{Du2024a}, along with its coefficients in terms of the microscopic parameters. The dispersion relation
\begin{equation}
    \omega^2 = \frac{\lambda}{2} \mu^2 l_B^4 k^4 \approx 0.15 \mu^2 l_B^4 k^4
\end{equation}
is independent of the interaction strength in this calculation. Note that in an experiment, we usually fix the number of particles and not the chemical potential, and in that case the dispersion relation will depend on the interaction strength. 

One can repeat the same calculations for a square lattice or any other configuration of vortices and that would describe the fluctuations on top of this putative metastable state. For the square lattice, we find that the small deformations of the form (\ref{eqn:deformation}) have \emph{negative} energy \changes{(\appref{app:C2})} and therefore it is unstable, not metastable.

We end this section by noting that the quadratic dispersion of the Tkachenko mode implies that there is no long-range order of $e^{i\hphi}$ in the 2+1D LLL superfluid, due to a generalized Hohenberg-Mermin-Wagner theorem~\cite{YuanChenYe2020,StahlLakeNandkishore2022,KapustinSpodyneiko2022,Spodyneiko2023}. Instead, there is quasi-long range order and the field theory for the Tkachenko mode describes the spin-wave type excitations similar to the 1+1D superfluid, ignoring topological defects. Although the Tkachenko modes also describe the phonon motion of the vortex lattice (note that translation is spontaneously broken unlike $U(1)$ charge conservation), we believe it is inappropriate to call it a Nambu-Goldstone boson. 


\section{Discussion}
\label{sec:5}
We have derived the non-commutative EFT of the LLL superfluid starting from a Hamiltonian for bosons restricted to the LLL, using a coherent state path integral approach. Although this work does not produce any qualitatively new results about this phase, it gives insight into the origin of the non-commutative fields which was lacking in previous works. Unlike the previous work~\cite{Du2024a} which justifies this non-commutative EFT using boson-vortex duality, this direct approach has the advantage of being able to make quantitative statements about the low-energy physics and comparison to the microscopic parameters of the model. 

The Weyl symbol produces an invertible map between a long-wavelength NC function and a long-wavelength commutative function, which tells us that we can use either language to describe the effective field theory. The utility of NC field theory is in the fact that it keeps the non-commutative magnetic translation symmetry manifest while retaining only the low-energy, long-wavelength degrees of freedom. In this sense, NC effective field theory is just regular Wilsonian EFT where the terms are organized appropriately in a symmetry covariant fashion.

A related aspect we want to emphasize is that this derivation suggests that the EFT for the LLL superfluid discussed here and in Ref.~\onlinecite{Du2024a} is not an NC field theory in the sense that it is used in older literature~\cite{DouglasNekrasov2001}. One of the hallmarks of NC field theory is the presence of UV-IR mixing~\cite{Minwalla2000}, where an IR length scale $L$ is related to a UV length scale $l_B^2 / L$. This feature often manifests in lowest Landau level theories such as HLR theory as a relationship between the momentum of a quasiparticle and its dipole moment~\cite{PasquierHaldane1998}. In contrast, the theory developed in this paper has a UV distance cutoff $\Lambda^{-1}$ that is greater than the length scale set by the non-commutativity parameter, and therefore it cannot capture the effects of such UV-IR mixing. Nevertheless, it is important to appreciate that in this context the NC field theory arises as a natural language to keep track of the symmetries for the EFT, and is unlike UV complete non-commutative field theory like the one suggested for the HLR state and its descendants~\cite{DongSenthil2020,Goldman2022}. On the formal side, it may be appropriate to think of two qualitatively different kinds of NC field theories, parametrized by the dimensionless ratio between the two characteristic length scales of the EFT -- the UV cutoff $\Lambda^{-1}$ and the non-commutativity parameter $l_B$.

Exploring the dynamical consequences of this NC field-theoretic description is an interesting direction for future work, particularly hydrodynamics in the presence of slowly varying trap potentials. A varying potential breaks magnetic translation symmetry and thus dipole conservation~\cite{Spodyneiko2023}, leading to chiral charge transport perpendicular to the potential gradient. A similar mode should also emerge at the boundary of rotating BECs~\cite{Jeevanesan2022}. A quantitative analysis of this mode within the NC effective field theory framework is left for future work.

Another open problem is the physics of vortex lines of $\hphi$ in space-time~\cite{NguyenMoroz2023,Du2024b}. Large vortex loops of the coarse-grained field $\hphi$ confined to a time slice correspond to cooperative ring exchange processes~\cite{KivelsonCRE1,KivelsonCRE2,LeeBaskaranKivelsonCRE} in the triangular vortex lattice (since the system lacks Lorentz invariance, space and time are not on equal footing). Reference~\onlinecite{GhoshBaskaranCRE} suggests, based on an approximate calculation using vortex coordinates, that the proliferation of cooperative ring exchange processes in the path integral would drive a transition to a bosonic fractional quantum Hall state. It would be valuable to reproduce their results using the effective field theory developed in this work.

We also note a related study on LLL superfluid hydrodynamics~\cite{MusserGoldmanSenthil2024}, which employs a different coherent-state path integral approach to describe rotating BECs. Unlike our work, which focuses on vortex lattices and their fluctuations, their approach describes experimentally relevant cases where the BEC condenses into a single angular momentum state. 

\section{Acknowledgments}
We thank Jason Alicea, Hart Goldman, Xiaoyang Huang, Dung Xuan Nguyen, Valerio Peri, Leo Radzihovsky, and T. Senthil for useful discussions and valuable comments. We also thank the organizers of the Prospects in theoretical physics (PiTP) 2024 program at IAS, Princeton for their hospitality where part of this work was done. This work was primarily led and supported by the U.S. Department of Energy, Office of Science, National Quantum Information Science Research Centers, Quantum Science Center. We appreciate the support received from the Institute of Quantum Information and Matter.

\bibliography{apssamp}

\onecolumngrid

\clearpage
\appendix
\section{Coarse grained wavefunction products}
\label{app:CGIP}
We show that the product 
\begin{equation}
    u_{\kk}(\rr) = \Phi_0^\ast(\rr) e^{i \kk \cdot\hr} (\Phi_0(\rr)) = \Phi_0^\ast(\rr) \Phi_{\kk}(\rr) \approx e^{i \kk \cdot \rr}.
\end{equation}
To show this, we look at the effect of lattice translation on $u_{\kk}(\rr)$, where the translation vector is the lattice vector of the reference state $\ket{\Phi_0}$. Let $\aa = (a_1,a_2)$ be a lattice translation vector. 
\begin{align}
    u_{\kk}(\rr-\aa) &= \Phi_0^\ast(\rr-\aa) \Phi_{\kk}(\rr-\aa) \\
    &= \braket{\Phi_0}{\rr-\aa} \braket{\rr-\aa}{\Phi_{\kk}} \\
    &= \bra{\Phi_0}\PLLL\ket{\rr-\aa} \bra{\rr-\aa}\PLLL\ket{\Phi_{\kk}} \\
    &= \bra{\Phi_0} e^{i \aa \wedge \rr l_B^{-2}} \hat{T}^\dagger_{\aa} \PLLL\ket{\rr} \bra{\rr} \PLLL \hat{T}_{\aa} e^{-i \aa \wedge \rr l_B^{-2}}\ket{\Phi_{\kk}} \\
    &= \bra{\Phi_0}  \hat{T}^\dagger_{\aa} \PLLL\ket{\rr} \bra{\rr} \PLLL \hat{T}_{\aa} \ket{\Phi_{\kk}} \\
    &= \bra{\Phi_0} \PLLL\ket{\rr} \bra{\rr}\PLLL \ket{\Phi_{\kk}} e^{-i \kk \cdot \aa} \\
    &= u(\rr) e^{-i \kk \cdot \aa}
\end{align}
This means that 
\begin{align}
    u_{\kk}(\rr) = e^{i \kk \cdot \rr}\term{u_0(\kk) + \sum_{\boldsymbol{K}\neq 0} u_{\boldsymbol{K}}(\kk) e^{i \boldsymbol{K}\cdot \rr}}
\end{align}
where the sum is over the reciprocal lattice vectors $\abs{\boldsymbol{K}} \sim l_B^{-1}$. When coarse graining, we can ignore all these terms. Moreover, we can argue by the reflection symmetry of the vortex lattice that $u_0(\kk) = 1 + \mathcal{O}(l_B^2 k^2)$, when $\ket{\Phi_0}$ is normalized appropriately. Therefore, we have shown that for small momenta $\kk$, the product of the wavefunctions satisfy 
\begin{equation}
    \Phi_0^\ast(\rr) e^{i \kk \cdot\hr} (\Phi_0(\rr)) = \Phi_0^\ast(\rr) \Phi_{\kk}(\rr) \approx e^{i \kk \cdot \rr}.
\end{equation}
upon coarse graining.

\section{Tools in non-commutative field theory}
\label{app:Tools}
We describe the useful mathematical tools in non-commutative field theory in this appendix. Most of this discussion can be found in standard reviews of non-commutative field theory, such as \cite{DouglasNekrasov2001}.

\subsection{Weyl map and Moyal products}
We motivated the Weyl map, also known as the Weyl symbol, in the main text. $W$ is a linear, invertible map between the vector space of NC functions and the vector space of functions on $\mathbb{R}^2$ (we will assume that the functions are \changes{well-behaved} without going into too much detail) given by 
\begin{equation}
    W[e^{i\kk\cdot \hr}] = e^{i\kk \cdot \rr} \qquad W[\hf](\rr) \equiv f(\rr) = \sum_{\kk} f_{\kk} e^{i \kk \cdot \rr}.
\end{equation}
What is the Weyl symbol of a product of two NC functions? We can show that 
\begin{align}
    W\comm{\hf \hg}(\rr) &= \sum_{\kk,\kk'} f_{\kk} g_{\kk'} W\comm{e^{i\kk\cdot \hr} e^{i\kk'\cdot \hr}} \\
    &= \sum_{\kk,\kk'} f_{\kk} g_{\kk'} W\comm{e^{i(\kk + \kk')\cdot \hr} e^{i\kk\wedge \kk' l_B^2/ 2}} \\
    &= \sum_{\kk,\kk'} f_{\kk} g_{\kk'} e^{i(\kk + \kk')\cdot \rr} e^{i\kk\wedge \kk' l_B^2/ 2} \\
    &= f(\rr) \exp^{-i \frac{l_B^2}{2} \overleftarrow{\partial} \wedge \overrightarrow{\partial}} g(\rr) \\
    &\equiv (f \star g)(\rr).
\end{align}
This is known as the Moyal product. The Moyal product is simply derived from a non-commutative matrix product, so it inherits all the familiar properties such as associativity and distributivity over additions, and it is non-commutative as are matrix products. It is also common to define a Moyal bracket which is related to the Weyl symbol of the commutator of two functions. 
\begin{align}
    W\comm{\hf \hg - \hg \hf}(\rr) &= (f \star g)(\rr) - (g \star f)(\rr) \\ 
    &= - 2 i f(\rr)\sin\term{\frac{l_B^2}{2} \overleftarrow{\partial} \wedge \overrightarrow{\partial}} g(\rr) \\
    &\equiv i \curly{\curly{f,g}}(\rr).
\end{align}
We want to emphasize that Weyl symbols and Moyal products are not fundamental to NC field theories -- they are simply an artifact of a useful map to regular fields. We can do all the calculations in NC field theory without referring to these objects by working with the NC functions. It is sometimes convenient to think in terms of the Weyl symbols and replacing all products by Moyal products.

\subsection{Derivatives and integrals}
Since we want to write down a field theory, it is useful to define analogs of derivatives and integration over the non-commutative space. We will motivate this using the plane-wave functions and using analogies to regular field theories. The derivatives for NC functions is a linear map defined as 
\begin{equation}
    \partial_j e^{i \kk \cdot \hr} = i k_j e^{i \kk \cdot \hr}.
\end{equation}
It is easy to show that this definition of a derivative satisfies the product rule $\partial_j(\hf \hg) = \partial_j(\hf) \hg + \hf \partial_j(\hg)$. The derivative is also compatible with the Weyl map
\begin{equation}
    W\comm{\partial_j \hf(\hr)} = \partial_j f(\rr)
\end{equation}
since we define it using the plane-wave functions. Because of this structure, we use the same notation for derivatives of NC fields and regular fields. There is a different representation of the derivative that's quite useful. We can show that 
\begin{equation}
    \partial_j \hf(\hr) = -i l_B^{-2} \epsilon_{jk} \comm{\hat{r}_k, \hf(\hr)}.
\end{equation}
One can also check that these derivatives commute, i.e. $\d_x\d_y \hf = \d_y \d_x \hf$.

We define the integral as a linear map from the space of NC functions to $\mathbb{C}$ given by
\begin{equation}
    \int_{\hr} e^{i \kk \cdot \hr} = (2 \pi)^2 \delta^{(2)}(\kk).
\end{equation}
In the operator language, this corresponds to taking the trace of the integrand. So we will often use the notation 
\begin{equation}
    \int_{\hr} \hf \equiv \Tr\comm{\hf} = \int \frac{d^2\kk}{(2\pi)^2} f_{\kk} (2 \pi)^2 \delta^{(2)}(\kk) = f_0 = \int d^2\rr f(\rr).
\end{equation}
Thinking of the integral as a trace will make certain manipulations manifest, such as the cyclicity of the trace. The integral of a product of NC functions is 
\begin{align}
    \Tr\comm{\hf \hg} &= \int \frac{d^2\kk}{(2\pi)^2} f_{\kk} g_{-\kk} \\
    &= \int d^2 \rr f(\rr) g(\rr).
\end{align}
The Moyal product is irrelevant for a quadratic integral. This is not true when you have cubic or higher powers of functions as you can have momentum conservation with non-trivial wedge products of the momentum vectors. The familiar expectation that the integral of a total derivative is zero is true for NC functions as well. Another way to understand it in this case is that the trace of a commutator is zero due to the cyclicity property. This means that we can do integration by parts for NC functions
\begin{equation}
    \Tr\comm{\partial_i \hf \hg} = -\Tr\comm{ \hf \partial_i \hg}.
\end{equation}

We can also do Gaussian functional (path) integrals of NC functions. We can look at a generic Gaussian integral where $A$ is a positive linear operator acting on the space of NC functions,
\begin{align}
    \int D\hf \exp\term{\Tr\comm{-\hf A \hf^\dagger + \hf \hg^\dagger  + \hg' \hf^\dagger }} &\equiv \int \prod_{\kk} \frac{df_{\kk} df^\ast_{\kk}}{2\pi i} \exp\term{\Tr\comm{-\hf A \hf^\dagger + \hf \hg^\dagger  + \hg' \hf^\dagger }} \\
    &= \det A^{-1} \exp\term{\Tr\comm{\hg' A^{-1} \hg^\dagger}}
\end{align}
This can be easily proved by using the Weyl map and arguing that quadratic integrals are agnostic toward the Moyal product -- we can treat them as regular functions and the standard formula in textbook many-body physics follows. Alternatively, we can note that $\Tr[\hf \hg^\dagger] = \sum_{ij} f_{ij} g^\ast_{ij}$ since $\hf,\hg$ are matrices in the LLL. So the trace of the product is like a dot product if you think of the matrices as vectors. Again, the standard formula for Gaussian integrals of vector inputs returns us the stated expression. 

We can also do a path integral for a real NC function (satisfying $\hf = \hf^\dagger$),
\begin{align}
    \int D\hf \exp\term{\Tr\comm{-\half \hf A \hf + \hf \hg }} &\equiv \int \prod_{\kk\in \half\text{BZ}} \frac{df_{\kk} df^\ast_{\kk}}{2\pi i} \exp\term{\Tr\comm{-\half \hf A \hf  + \hf \hg }} \\
    &= \det A^{-1/2} \exp\term{\Tr\comm{\half \hg A^{-1} \hg}}
\end{align}
Note that these Gaussian integrals are useful in performing Hubbard-Stratonovich transformations in NC field theory. Like quantum field theory, Hubbard-Stratonovich fields are often useful to be introduced as the correct classical variables describing symmetry-broken states. One can also write down closed-form expressions of expectation values of trace of field bilinears and a generalization of Wick's theorem for NC fields, like usual QFT. We can in principle use this to perform perturbative calculations starting from this free-field theory.

\subsection{Spatial symmetries}
We finish this appendix by describing spatial symmetries such as translation and rotation in this framework. Like we emphasized in the main text, the utility of NC field theory is that these spatial symmetries are manifest in this language and serves as an aid to write effective field theories.

In this section, we always define $\aa = l_B^2(\wedge \kk)$. Let us first look at translations of NC fields. We can show that for any NC field $\hf$,
\begin{align}
    \hf(\hr-\aa) = e^{i\kk\cdot \hr} \hf(\hr) e^{-i\kk\cdot \hr}.
\end{align}
Generally, if an object transforms as $\hf(\hr) \mapsto e^{i\kk\cdot \hr} \hf(\hr) e^{-i\kk\cdot \hr}$ under translation by $\aa$ we say it transforms covariantly under translations. This structure adds and multiplies well -- if $\hf,\hg$ are covariant then their products and sums are covariant as well. Importantly, the field operator that we perform the path integral over is not covariant. We saw in the main text that a symmetry acts on a single particle state as
\begin{equation}
    \ket{\hf} \mapsto \ket{\hat{U}\hf},
\end{equation}
which means that a field operator transforms as 
\begin{equation}
    \hf \mapsto e^{i \kk \cdot \hr}\hf, \qquad \hf^\dagger \mapsto \hf^\dagger e^{-i \kk \cdot \hr},
\end{equation}
under translation by $\aa$. This is because the field operator is a charged object, and in the LLL there is a nontrivial interplay between charge and spatial symmetries. We can construct uncharged operators which transform covariantly, such as 
$\hf \hf^\dagger$, this is analogous to the density operator. Note the nontrivial ordering -- although $\hf^\dagger \hf$ is also uncharged, it does not transform covariantly under spatial symmetries -- it actually remains invariant. 

Because of its nice properties, it is useful to construct the Lagrangian out of covariant objects $\mathcal{L}(\hf \hf^\dagger, \partial_i(\hf\hf^\dagger),\dots)$ and we are guaranteed that the action will be symmetric. Note that this is not a necessary condition to have a symmetric theory, we can have nontrivial terms in the Lagrangian which are invariant under symmetries up to a total derivative which does not change the physics (an example of such a term in regular field theory is the Chern-Simons term which is not invariant under gauge symmetries).

The derivatives of charged operators transform nontrivially under translations. We can show that 
\begin{align}
    \partial_j(\hf(\hr)) &\mapsto \partial_j(e^{i\kk\cdot \hr}\hf(\hr)) \\
    &= e^{i\kk\cdot \hr} \term{\partial_j + i k_j}\hf(\hr).
\end{align}
Similarly, 
\begin{align}
    \partial_j(\hf^\dagger(\hr)) &\mapsto \partial_j(\hf^\dagger(\hr)e^{-i\kk\cdot \hr}) \\
    &=  \term{\partial_j - i k_j}\hf^\dagger(\hr) e^{-i\kk\cdot \hr}.
\end{align}
On the other hand, the derivative acting on an uncharged operator transforms trivially. 
\begin{align}
    \partial_j(\hg(\hr)) &\mapsto \partial_j(e^{i\kk\cdot \hr}\hg(\hr)e^{-i\kk\cdot \hr}) \\
    &= e^{i\kk\cdot \hr} {\partial_j}\hg(\hr) e^{-i\kk\cdot \hr}.
\end{align}

The action of rotations is slightly more tedious to work with. One can show that the rotation operator in the LLL is $e^{i \omega \hr^2 /2}$, which rotates the one-particle wavefunction by an angle $\Omega \equiv \omega l_B^2$. This can be easily verified by checking the commutation relations between the generators of magnetic translation and this operator. 

An NC field transforms covariantly under rotations as
\begin{equation}
    \hf(\hr) \mapsto \hf(U^{-1}(\Omega)\hr) = e^{i \frac{\omega}{2} \hr^2} \hf(\hr) e^{-i \frac{\omega}{2} \hr^2}, \qquad U(\Omega) = \begin{pmatrix} \cos\Omega & -\sin\Omega \\ \sin\Omega & \cos\Omega \end{pmatrix}.
\end{equation}
Under an infinitesimal rotation, it transforms as 
\begin{equation}
    \hf(\hr) \mapsto \hf(\hr) + i \frac{\omega}{2}\comm{\hr^2,\hf(\hr)}.
\end{equation}
A charged (field) operator transforms as (this follows from the transformation law for a single particle quantum state)
\begin{equation}
    \hf(\hr) \mapsto e^{i \frac{\omega}{2} \hr^2} \hf(\hr).
\end{equation}
Again we see that the previously mentioned uncharged covariant operators ($\hf \hf^\dagger$) also transform covariantly under rotations. It is useful to see how derivatives of charged operators transform under the rotation symmetry. We find that, under a rotation
\begin{align}
    \partial_i\hf(\hr) &\mapsto \partial_i\term{e^{i \frac{\omega}{2} \hr^2}  \hf(\hr)} \\
    &= -il_B^2 \epsilon_{ij}\comm{\hat{r}_j,\term{e^{i \frac{\omega}{2} \hr^2}  \hf(\hr)}} \\
    &= -il_B^2 \epsilon_{ij} e^{i \frac{\omega}{2} \hr^2} \term{ U(\Omega)_{jj'} \hat{r}_{j'} \hf(\hr) - \hf(\hr) \hat{r}_j} \\
    &= e^{i \frac{\omega}{2} \hr^2} \comm{U(\Omega)_{ii'}\frac{\partial}{\partial \hat{r}_{i'}}  \hf(\hr) + \hf(\hr)\term{-il_B^2 \epsilon_{ij}}\term{(U(\Omega)\hr)_j - \hat{r}_j }} \\
    &\approx e^{i \frac{\omega}{2} \hr^2} \comm{U(\Omega)_{ii'}\partial_{i'}  \hf(\hr) + \hf(\hr)\term{il_B^2 \epsilon_{ij}}\Omega \epsilon_{jj'}\hat{r}_{j'}} \\
    &=e^{i \frac{\omega}{2} \hr^2} \comm{U(\Omega)_{ii'}\partial_{i'}  \hf(\hr) - il_B^2 \Omega \hf(\hr)  \hat{r}_{i}}
\end{align}
where in the last two steps we make an approximation of small rotation angle. We find that the derivative transforms covariantly (indicated by the rotation of the derivative), but it also picks up an additional term where the field gets right multiplied by a position operator. Note that in the second step, we use the form of the derivative in terms of a commutator with the position operator. Similarly, the opposite charged field transforms under infinitesimal rotations as
\begin{equation}
    \partial_i\hf^\dagger(\hr) \mapsto  \comm{U(\Omega)_{ii'}\partial_{i'}  \hf^\dagger(\hr) + il_B^2 \Omega \hat{r}_{i} \hf^\dagger(\hr) }e^{-i \frac{\omega}{2} \hr^2}
\end{equation}
One can also show that the derivative acting on an uncharged operator acts covariantly, i.e.
\begin{equation}
    \partial_i\hg(\hr) \mapsto  e^{i \frac{\omega}{2} \hr^2}\comm{U(\Omega)_{ii'}\partial_{i'}  \hg(\hr)}e^{-i \frac{\omega}{2} \hr^2}
\end{equation}

\subsection{Polar Decomposition}
We end by noting that any field operator $\hf$ can be written in the form $\hu \sqrt{\hat{n}}$ using polar decomposition, where $\hu$ is a unitary matrix and $\hat{n}$ is a Hermitian positive semi-definite matrix acting on the LLL. In this language, under a symmetry $\hU$ the individual parts transform as 
\begin{equation}
    \hu \mapsto \hU\hu, \qquad \hn \mapsto \hn.
\end{equation}
Each individual part of the polar decomposition remains unitary and positive Hermitian respectively, under the symmetry transformation. This decomposition is quite helpful to understand LLL superfluidity, as we discuss in the main text.

\section{Numerical estimation of effective field theory parameters}\label{app:C}
In this Appendix, we discuss the numerical calculation of the effective parameters described in the main text. The idea is to match the energy of a LLL coherent state in the microscopic Hamiltonian (a numerical calculation) and match that with the energy calculated from the effective field theory. 

\subsection{Estimating $g^\ast$}
\label{app:C1}
To calculate $g^\ast$, we look at the reference state $\Ket{\Phi_0}$ and its energy in the microscopic action \eqnref{eqn:microscopic_action}. The energy will be given by 
\begin{align}
    E_0 &= \int d^2\rr \comm{-\mu \braket{\Phi_0}{\rr}\braket{\rr}{\Phi_0} + \frac{g}{2}  \braket{\Phi_0}{\rr}\braket{\rr}{\Phi_0}\braket{\Phi_0}{\rr}\braket{\rr}{\Phi_0}} \\
    &= \int d^2\rr \comm{-\mu \abs{\Phi_0(\rr)}^2 + \frac{g}{2} \abs{\Phi_0(\rr)}^4}.
\end{align}
As $\abs{\Phi_0(\rr)}^2$ is periodic, we can integrate this over one unit cell. We will normalize $\Phi_0(\rr)$ such that $\int_{\text{Unit Cell}} d^2 \rr \abs{\Phi_0(\rr)}^2 = 2\pi l_B^2$, the area of the unit cell. With this convention, we find that 
\begin{equation}
    \int_{\text{Unit Cell}} d^2 \rr \abs{\Phi_0(\rr)}^4 \approx 1.160(2\pi l_B^2)
\end{equation}
using a straightforward numerical integration. So the total energy in the microscopic description becomes
\begin{equation}
    E_0 \approx \term{-\mu + 1.160 \frac{g}{2}} \text{ per unit area}.
\end{equation}
Note that a more efficient computation of the factor $1.160$ is described in \cite{YangMacdonald2004}. 

Now, let us go to the effective description in terms of non-commutative functions. The state $\Ket{\Phi_0}$ is described by the trivial function $\hf = 1$. The energy is just 
\begin{align}
    E_0 &= \Tr \comm{-\mu \hf \hf^\dagger + \frac{g^\ast}{2} \hf \hf^\dagger\hf \hf^\dagger},
\end{align}
plus higher-order terms which we will ignore. Using the Weyl symbols, this expression can be written as 
\begin{align}
    E_0 &= \int d^2\rr \comm{-\mu f \star f^\ast + \frac{g^\ast}{2} f \star f^\ast \star f \star f^\ast}.
\end{align}
Since for our reference state $\hf =1$, we have $f(\rr) = 1$ and therefore the energy is simply
\begin{equation}
    E_0 = \term{-\mu +  \frac{g^\ast}{2}} \text{ per unit area}.
\end{equation}
Matching the microscopic and effective theories give us $g^\ast \approx 1.160 g$. One can do an analogous calculation for the square lattice to find $g^\ast \approx 1.180$, indicating that the square lattice is less energetically favorable compared to the triangular lattice. 

\subsection{Estimating $\lambda$}
\label{app:C2}
\begin{figure}
    \centering
    \includegraphics[width=0.98\linewidth]{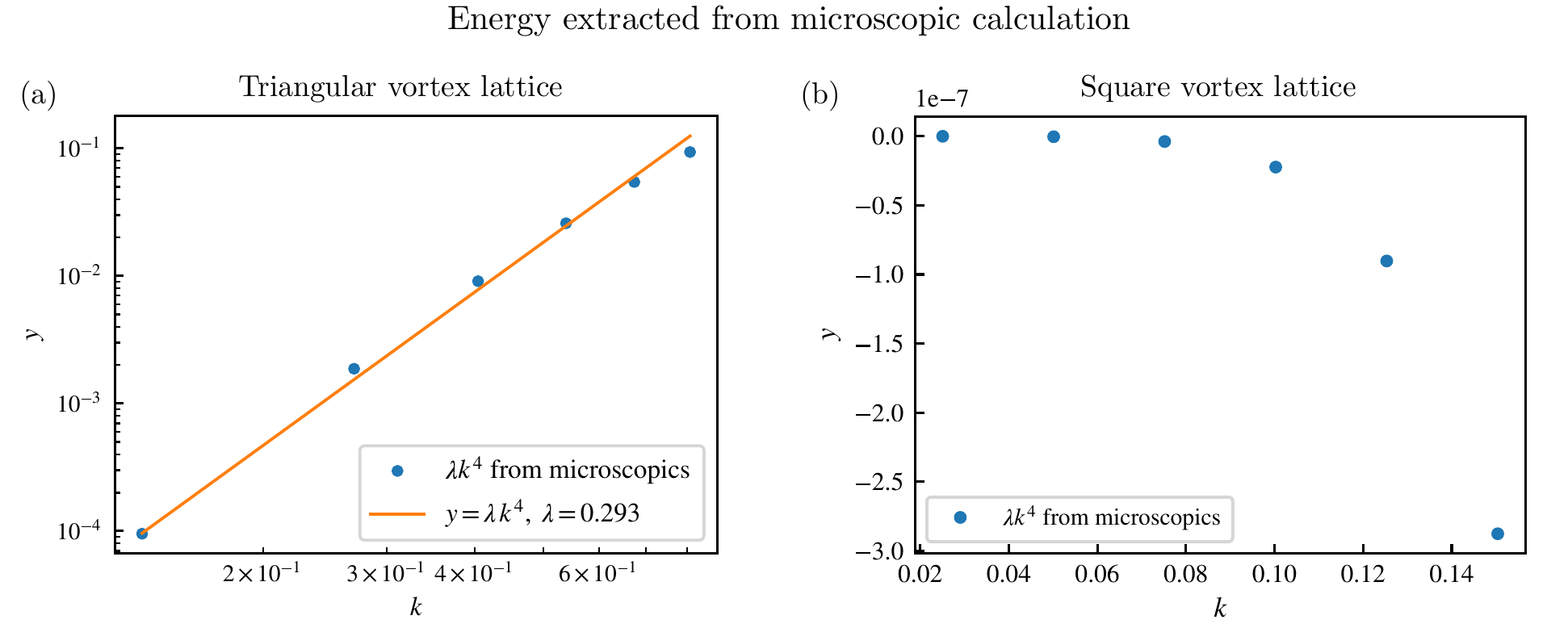}
    \caption{Extracting the EFT coefficient $\lambda$ from numerics using $\varphi_0 = 0.1$.}
    \label{fig:2}
\end{figure}
To estimate $\lambda$, we consider a typical ``oscillating'' configuration of vortices, given by $\hu = e^{i\hphi} =  \exp(i \varphi_0 \cos(\kk\cdot \hr))$, which corresponds to $\hf = \sqrt{n_0} \exp(i \varphi_0 \cos(\kk\cdot \hr))$. We calculate the interaction energy for this state $\Ket{\hf}$ in the microscopic action \eqnref{eqn:microscopic_action}, relative to the ground state $\hf = \sqrt{n_0}$, $n_0 = \mu / g^\ast$. That is, we compute
\begin{equation}
    \int d^2 \rr \frac{g}{2}\comm{\abs{\Phi_{\hf}(\rr)}^4 - \abs{\Phi_{0}(\rr)}^4 } 
    \label{eqn:numerical_computation}
\end{equation}
and match it with the energy from the effective action 
\begin{align}
    \Tr\comm{\frac{\mu^2 \lambda l_B^4}{2 g^\ast} (D_{ij} \hphi)^2} &= \int d^2 \rr \comm{\frac{\mu^2 \lambda l_B^4}{2 g^\ast} (D_{ij} W[\hphi])^2} 
\end{align}
To extract the value of $\lambda$. Numerically, we use $\varphi_0 = 0.1,0.2$, and $\kk = (0,k_y)$ to simplify the computation, and we use the numerically obtained value $g^\ast / g \approx 1.16$ for the triangular lattice ($1.18$ for square lattice). Simplifying the expression from the EFT, 
\begin{align}
    \int d^2 \rr \comm{\frac{\mu^2 \lambda l_B^4}{2 g^\ast} (D_{ij} W[\varphi_0 \cos(k_y \hy)])^2}  &= \int d^2 \rr \comm{\frac{\mu^2 \lambda l_B^4}{2 g^\ast} (D_{ij} (\varphi_0 \cos(k_y y))^2}\\
    &= \int d^2 \rr \comm{\frac{\mu^2 \lambda l_B^4}{2 g^\ast} \term{\half\varphi_0^2 k_y^4 \cos^2(k_y y)}} \\
    &= \text{Area} \times \comm{\frac{\mu^2 \lambda l_B^4}{2 g^\ast} \term{\frac{1}{4}\varphi_0^2 k_y^4 }}. \label{eqn:EFT_computation}
\end{align}

For the microscopic calculation, we compute \eqnref{eqn:numerical_computation} for discrete values of $k_y$ such that we can numerically integrate the expression over one Bravais cell of the residual discrete translation symmetry. Note that this becomes relatively easy as the strip-wavefunctions we use to construct the reference state $\ket{\Phi_0}$ (cf. eqns. (\ref{eqn:Phi_square},\ref{eqn:Phi_triangular})) are eigenstates of $\hy$. Matching with the EFT expression (\ref{eqn:EFT_computation}), we can get rid of all the dimensionful constants and use the known $g^\ast/g$ values to plot $\lambda k_y^4$ as extracted from the numerics. The results are plotted in \figref{fig:2} (the case of $\varphi_0 = 0.2$ looks quantitatively similar). For the triangular vortex lattice, we find $\lambda \approx 0.29$. Surprisingly, we find that the energy of the deformed configuration is lower than the reference configuration in the case of the square lattice, indicating that the square vortex lattice is unstable to fluctuations.

\end{document}